\PassOptionsToPackage{table,xcdraw}{xcolor}
\documentclass[sigconf,screen]{acmart}
\usepackage{listings}
\usepackage{booktabs}
\usepackage{xcolor}
\usepackage[most]{tcolorbox}
\usepackage[normalem]{ulem}

\usepackage{enumitem}

\usepackage{caption}
\AtBeginDocument{%
  }

\lstdefinelanguage{Python}{
    morekeywords={class, def, return, try, except, raise, from},
    keywordstyle=\color{blue},
    stringstyle=\color{green},
    commentstyle=\color{gray},
    morecomment=[l]{\#},
}

\lstdefinelanguage{diff}{
    morecomment=[f][\color{green}]{+},
    morecomment=[f][\color{red}]{-},
    morecomment=[f][\color{blue}]{@@},
}

\lstdefinelanguage{errorlog}{
    morecomment=[f][\color{red}]{E},
    morecomment=[f][\color{magenta}]{?},
}

\usepackage{algorithm}
\usepackage{algorithmic}

\usepackage{listings}
\usepackage{xspace}

\usepackage[T1]{fontenc}

\usepackage[utf8]{inputenc}

\usepackage{microtype}

\usepackage{graphicx}

\usepackage{subcaption}

\usepackage{soul}
\usepackage{framed}
\usepackage{multirow}
\usepackage{siunitx}

\newcommand*\colourcheck[1]{%
	\expandafter\newcommand\csname #1check\endcsname{\textcolor{#1}{\ding{52}}}%
}
\colourcheck{blue}
\colourcheck{green}
\colourcheck{red}

\definecolor{custom-blue}{rgb}{0,0,0}

\newcommand{\tool}{\textsc{EvolRepair}\xspace}

\newboolean{showcomments}
\setboolean{showcomments}{true}
\ifthenelse{\boolean{showcomments}}
 { \newcommand{\mynote}[2]{
      \fbox{\bfseries\sffamily\scriptsize#1}
        {\small$\blacktriangleright$\textsf{\emph{#2}}$\blacktriangleleft$}}}
        { \newcommand{\mynote}[2]{}}

\newcolumntype{L}[1]{>{\raggedright\arraybackslash}p{#1}}

\newcommand{\code}[1]{{\small\texttt{#1}}}
\usepackage{amsthm}
\definecolor{dkgreen}{rgb}{0,0.6,0}
\definecolor{gray}{rgb}{0.5,0.5,0.5}
\definecolor{lightgray}{rgb}{211, 211, 211}
\definecolor{mauve}{rgb}{0.58,0,0.82}

\definecolor{custom-red}{rgb}{1,0,0}
\definecolor{custom-blue}{rgb}{0,0,1}

\definecolor{c1}{HTML}{f4cccc}
\definecolor{c2}{HTML}{f5cdcd}
\definecolor{c3}{HTML}{fffcfc}
\definecolor{c4}{HTML}{ffffff}
\definecolor{c5}{HTML}{ffffff}
\definecolor{c6}{HTML}{fffdfd}
\definecolor{c7}{HTML}{f5cfcf}
\definecolor{c8}{HTML}{fffbfb}
\definecolor{c9}{HTML}{ffffff}
\definecolor{c10}{HTML}{fffdfd}
\definecolor{c11}{HTML}{fefafa}
\definecolor{c12}{HTML}{fef7f7}
\definecolor{c13}{HTML}{ffffff}
\definecolor{c14}{HTML}{fffefe}
\definecolor{c15}{HTML}{ffffff}
\definecolor{c16}{HTML}{fefafa}
\definecolor{c17}{HTML}{fdf3f3}
\definecolor{c18}{HTML}{fffefe}
\definecolor{c19}{HTML}{fdf5f5}
\definecolor{c20}{HTML}{ffffff}

\begin{document}

\title{Semantic Evolution over Populations for LLM-Guided Automated Program Repair}

\author{Cuong Chi Le}
\orcid{0009-0006-1322-7396}
\affiliation{%
 \institution{University of Texas at Dallas}
 \city{Dallas}
 \state{Texas}
 \country{USA}
 \postcode{75080-3021}
}

\author{Minh Le-Anh}
\orcid{0009-0000-2859-7038}
\affiliation{%
 \institution{FPT Software AI Center}
 \state{Hanoi}
 \country{Vietnam}
}

\author{Cuong Duc Van}
\orcid{0009-0008-6700-1477}
\affiliation{%
 \institution{Hanoi University of Science and Technology}
 \state{Hanoi}
 \country{Vietnam}
}

\author{Tien N. Nguyen}
\orcid{0009-0006-7962-6090}
\affiliation{%
 \institution{University of Texas at Dallas}
 \city{Dallas}
 \state{Texas}
 \country{USA}
 \postcode{75080-3021}
}



\setcopyright{none}

\settopmatter{printacmref=false, printfolios=false}

\renewcommand\footnotetextcopyrightpermission[1]{} 

\begin{abstract}
Large language models (LLMs) have recently shown strong potential for automated program repair (APR), particularly through iterative refinement that generates and improves candidate patches. However, state-of-the-art iterative refinement LLM-based APR approaches cannot fully address challenges 
including maintaining useful diversity among repair hypotheses, identifying semantically related repair families, composing complementary partial fixes, exploiting structured failure information, and escaping structurally flawed search regions. In this paper, we propose \textbf{Population-Based Semantic Evolution framework} for {\bf APR iterative refinement}, called \tool, that formulates LLM-based APR as a semantic evolutionary algorithm. \tool reformulates the search paradigm of classic genetic algorithm for APR, but replaces its syntax-based operators with semantics-aware components powered by LLMs and structured execution feedback. Candidate repairs are organized into behaviorally coherent groups, enabling the algorithm to preserve diversity, reason over repair families, and synthesize stronger candidates by recombining complementary repair insights across the population. By leveraging structured failure patterns to guide search direction, \tool can both refine promising repair strategies and shift toward alternative abstractions when necessary. Our experiments show that \tool substantially improves repair effectiveness over existing LLM-based APR approaches.
\end{abstract}







\maketitle

\section{Introduction}
\label{sec:intro}

Automated Program Repair (APR) has evolved along several distinct
methodological trajectories over the past decade. Early systems
primarily followed the test-suite--based generate-and-validate
paradigm, where candidate patches are enumerated and validated against
a test suite serving as the correctness oracle. Search-based repair
methods such as \textsc{GenProg}~\cite{weimer2009genprog,le2011genprog}
operationalized this idea via genetic programming over syntactic
edits, using evolutionary mutation and selection guided by test
outcomes. Template-driven approaches (e.g., \textsc{SPR}~\cite{long2015spr}) further structured the search space by applying parameterized transformations at suspicious program locations. Learning-guided ranking models (e.g., \textsc{Prophet}~\cite{long2016prophet}) subsequently improved patch prioritization by statistically modeling properties of human-written fixes. While effective within constrained edit spaces, these systems often struggle with patch overfitting and limited semantic generalization beyond the test~suite.

Moving beyond syntactic search, semantics- and constraint-based
repair approaches introduced symbolic reasoning and synthesis. Systems
such as \textsc{SemFix}~\cite{nguyen2013semfix} and \textsc{Nopol}~\cite{xuan2017nopol} combined~symbolic execution with constraint solving to synthesize repairs consistent
with test cases. \textsc{Angelix}~\cite{mechtaev2016angelix}
improved scalability by leveraging angelic reasoning over execution
traces to synthesize patches. This direction offers strong semantic guarantees but often face scalability challenges and sensitivity to constraint encoding, and may still suffer from overfitting when test oracles are incomplete~\cite{le2018overfitting}. 

A complementary direction leverages the redundancy of historical bug
fixes through mining examples and fixing patterns (e.g., \textsc{PAR}~\cite{kim2013par}). \textsc{FixMiner}~\cite{koyuncu2018fixminer} extracts fine-grained atomic fix patterns from large patch corpora to guide repair
pipelines. This line of work bridges manually curated knowledge and
automated synthesis by exploiting recurring structural regularities in
real-world patches.

With the rise of neural networks, deep learning-based APR reframed repair
as a translation task from buggy to fixed code. Sequence-to-sequence
models such as \textsc{Sequencer}~\cite{chen2019sequencer} learned
patch generation directly from large bug-fix datasets. Later systems,
(e.g., \textsc{CoCoNuT}~\cite{lutellier2020coconut}, DLFix~\cite{icse20-dlfix}, DEAR~\cite{icse22-dear}) incorporated
contexts and ensemble strategies to improve
robustness. Although this direction significantly expanded the
repair search space beyond handcrafted templates, they remained
constrained by dataset coverage.

Recent advances in large language models (LLMs) have further
transformed APR. Modern LLM-based repair systems treat patch
generation as conditional code generation with powerful pretrained
priors. Empirical studies demonstrate that LLMs can outperform many
traditional APR systems under appropriate prompting and validation
workflows~\cite{xia2023plm_apr,xia2024automated,fan2023apr_llm_outputs}. 
To further improve fix generation quality, retrieval-augmented approaches incorporate similar bug-fix exemplars into the context (e.g., RAP-Gen~\cite{wang2023rapgen}). Moreover,
agentic APR frameworks (e.g., \textsc{AutoCodeRover}~\cite{zhang2024autocoderover}, \textsc{RepairAgent}~\cite{bouzenia2025repairagent}) introduce explicit
planning and tool use, allowing the model to iteratively localize
faults, inspect context, invoke compilation and testing tools, and
update its repair strategy dynamically.

Approaches like REx~\cite{rex2024neurips} improves APR for LLMs via {\em iterative refinement}. It formulates iterative code repair as a budgeted search problem: given a limited refinement budget and a set of candidate repair trajectories, the APR model decides which trajectory to refine next with LLM. Using Thompson Sampling, REx explicitly balances exploration and exploitation when selecting the next candidate, maintaining posterior beliefs over each candidate's repair potential. 

In this work, we focus on {\em iterative LLM-based refinement} methods, as they enable APR systems to explore a large repair space. They must satisfy the following.
\underline{First}, an effective approach must maintain {\em useful diversity among repair hypotheses}. Greedy refinement strategies often converge quickly to a narrow structural hypothesis, repeatedly producing minor variations of the same patch pattern while discarding alternative repair directions. Preserving diverse candidates that represent different repair ideas allows the search to explore multiple semantic possibilities in parallel and reduces the risk of premature convergence. \underline{Second}, the algorithm should be able to identify families of semantically related repairs. Many candidates differ syntactically yet correspond to the same underlying repair strategy, such as modifying a loop bound, inserting a guard condition, or adjusting a numerical constraint. Recognizing these relationships enables reasoning about repair strategies at a higher level and avoids repeatedly exploring near-duplicate candidates that represent the same hypothesis, or missing correct hypotheses.

\underline{Third}, effective search should support {\em combining partial repair insights} discovered across different candidates. In practice, the correct repair often requires multiple complementary changes that are rarely discovered simultaneously in a single generation step; different candidates may independently capture distinct aspects of the fix, and the ability to integrate these partial insights can substantially accelerate convergence toward a correct repair. \underline{Fourth}, the search process should exploit the {\em richer structure of failure information} rather than relying solely on scalar success signals such as the number of tests passed. \underline{Finally}, the APR algorithm must be able to adjust the abstraction level of the search when a region of the repair space proves fundamentally incorrect. Iterative refinement assumes that a candidate is close to the correct solution, but when the underlying repair hypothesis is flawed, continued local refinements only produce superficial variations of the same incorrect idea. Recognizing those and shifting the search toward alternative repair abstractions is essential to escape~structural~local~optima.

To address these requirements, we propose {\bf Population-Based Semantic Evolution} framework for {\bf APR iterative refinement}, called {\tool}, that formulates LLM-based APR as semantic evolutionary algorithm (EA). \tool inherits the foundational insight of classic EA such as GenProg~\cite{weimer2009genprog} that
maintaining a diverse population and applying selection pressure under
a test oracle enables robust search, but {\em replaces each core component
with stronger, semantics-aware counterparts}.  While GenProg relies on
largely {\em syntax-based mutation and crossover operators}, \tool treats the
LLM as a {\bf semantic mutation operator}, generating conditional
edits grounded in program context and observed failures rather than
applying blind syntactic perturbations. Candidate repairs are organized into behaviorally coherent {\bf groups} based on their execution outcomes, allowing to reason about {\em families of repair hypotheses} and preserve useful diversity across the population. 

Furthermore, {\tool} introduces {\bf population-level recombination beyond
pairwise crossover}. This population-level structure enables it to identify complementary partial repairs and synthesize stronger candidates by integrating insights discovered in different trajectories. Moreover, \tool leverages the structured patterns of test failures to guide the search direction, enabling the algorithm to determine whether a repair family should be refined further or whether the search should shift toward alternative repair abstractions. In this way,
{\tool} elevates the search process from individual trajectories to behaviorally structured populations, enabling the balance between exploration and exploitation, as well as the composition of repair strategies across the population.

\noindent{\bf Contributions.} This paper makes the following contributions:

{\bf 1. Population-Based Semantic Evolution for LLM-based APR:} 
We elevate repair from candidate-level refinement to population-level exploration, enabling the algorithm to maintain diverse repair hypotheses, reason over families of semantically related candidates, and compose complementary partial repairs across trajectories.

{\bf 2. Semantics-aware evolutionary operators for repair search.}
\tool revisits the population-based search paradigm of classic genetic repair systems (e.g., GenProg) and replaces their syntax-driven operators with stronger semantics-aware components.

{\bf 3. Empirical evaluation.} Our results show that {\tool} improves repair effectiveness over existing LLM APR approaches.
\section{Motivation}
\label{sec:motiv}


A recurring failure mode of LLM-based APR is that repair quality improves quickly on the \emph{dominant} execution paths of a program, while the candidate fix still fails on a small number of semantically difficult test cases. This is especially problematic for refinement-based APR methods: once a candidate patch already passes most test cases, subsequent refinement becomes biased toward small edits around that high-fitness candidate, because the feedback only suggests that the current direction is broadly promising. As a result, the search keeps optimizing within the same repair family, even when the remaining failures require a different repair abstraction, leading to a local optimum. Moreover, the globally correct repair is often not absent from the search space but \emph{distributed across multiple candidates}: different patches may independently repair different subsets of the failing behaviors, yet current methods typically continue improving only one candidate at a time. This prevents them from preserving diverse repair hypotheses, recognizing complementary partial fixes, and combining them into a stronger patch, which directly motivates our approach.

{\em A motivating example.} Consider the example in Figure~\ref{fig:motiv-rotated-search}, adapted from the well-known rotated-array binary search problem. The function should determine whether a target exists in a rotated sorted array that \emph{may contain duplicates}. The code contains several distinct control-flow strategies: the target may be found immediately; the left half may appear sorted; the right half may appear sorted; or duplicates may make the sorted-side inference ambiguous.

\lstdefinestyle{paperpython}{
  language=Python,
  basicstyle=\footnotesize\ttfamily,
  columns=fullflexible,
  keepspaces=true,
  frame=single,
  showstringspaces=false,
  xleftmargin=0.4em,
  xrightmargin=0.4em
}
\begin{figure}[t]
\begin{lstlisting}[language=Python, basicstyle=\footnotesize\ttfamily]
def search(nums: list[int], target: int) -> bool:
    left, right = 0, len(nums) - 1

    while left <= right:
        mid = (left + right) // 2

        if nums[mid] == target:
            return True

        if nums[left] <= nums[mid]:  # assume left half is sorted
            if nums[left] <= target < nums[mid]:
                right = mid - 1
            else:
                left = mid + 1
        else:  # assume right half is sorted
            if nums[mid] < target <= nums[right]:
                left = mid + 1
            else:
                right = mid - 1

        # missing case:
        # if nums[left] == nums[mid] == nums[right],
        # duplicates make the sorted-half inference ambiguous

    return False
\end{lstlisting}
\vspace{-15pt}
\caption{A motivating APR example based on rotated-array binary search with duplicates. The easy paths follow standard binary-search reasoning, but the duplicate-ambiguity path requires switching to a different repair abstraction rather than making only local comparison-level edits.}
\label{fig:motiv-rotated-search}
\end{figure}

This implementation is correct for the no-duplicate version of the problem, and therefore already passes many test cases. However, it is still incorrect for arrays with repeated values. In particular, when $\code{nums[left] == nums[mid] == nums[right]}$, the usual binary-search reasoning about which side is sorted becomes ambiguous. The hard test cases are therefore not just boundary cases; they correspond to a \emph{different semantic regime} of the algorithm.

Table~\ref{tab:motivation-tests} shows a small path-covering test suite. Most test cases exercise the standard binary-search behavior and are relatively easy to repair. The hard cases are the duplicate-ambiguity paths.

\begin{table}[t]
\centering
\footnotesize
\caption{Path-covering test cases for the motivating example.}
\label{tab:motivation-tests}
\vspace{-9pt}
\begin{tabular}{p{0.35\columnwidth}p{0.3\columnwidth}p{0.14\columnwidth}}
\toprule
\textbf{Covered path} & \textbf{Test input} & \textbf{Expected} \\
\midrule
Target found in rotated array & \texttt{([4,5,6,7,0,1,2], 0)} & \texttt{True} \\
Target absent & \texttt{([4,5,6,7,0,1,2], 3)} & \texttt{False} \\
Left side selected & \texttt{([6,7,0,1,2,4,5], 7)} & \texttt{True} \\
Right side selected & \texttt{([6,7,0,1,2,4,5], 4)} & \texttt{True} \\
Duplicate ambiguity I & \texttt{([1,1,1,3,1], 3)} & \texttt{True} \\
Duplicate ambiguity II & \texttt{([1,0,1,1,1], 0)} & \texttt{True} \\
\bottomrule
\end{tabular}
\end{table}

{\em Why current refinement methods get stuck.}
This example captures a failure pattern we repeatedly observed with state-of-the-art prompting approaches such as REx~\cite{rex2024neurips} and ChatRepair~\cite{xia2024automated}. A one-shot prompt often recognizes that the bug is in the search logic and proposes a \emph{comparison-level} fix, e.g., by adjusting `$\texttt{<}$' vs.\ `$\texttt{<=}$' or by slightly changing branch conditions. Such a patch may improve ordinary tests, but it usually still fails the duplicate-ambiguity cases.

LLM-based iterative refinement methods then continue from that already-strong candidate. Because the current patch passes most tests, the feedback encourages more local changes around the same binary-search abstraction: changing interval comparisons, shifting midpoint logic, or skipping one duplicated boundary. These refinements can increase fitness further, but they often remain trapped in the same hypothesis family: they still assume that one side can be reliably identified as sorted. Importantly, different candidate patches may already capture \emph{different correct partial behaviors} of the final repair, but single-trajectory refinement has no mechanism to preserve and recombine them. As a result, the search repeatedly improves \emph{within} an incorrect abstraction instead of composing the missing ambiguity-resolution step into a globally correct repair.

Table~\ref{tab:motivation-trajectory} illustrates a representative repair trajectory. The key issue is that each candidate is locally plausible and often passes more tests than the previous one, yet the search still fails to jump to the globally correct repair for all test cases.

\begin{table}[t]
\centering
\footnotesize
\tabcolsep 3pt
\caption{Representative repair trajectory on Figure~\ref{fig:motiv-rotated-search}.}
\label{tab:motivation-trajectory}
\vspace{-9pt}
\begin{tabular}{p{0.18\columnwidth}p{0.45\columnwidth}p{0.32\columnwidth}}
\toprule
\textbf{Patch family} & \textbf{Representative idea} & \textbf{Still fails} \\
\midrule
Buggy program & Standard rotated binary search & Duplicate ambiguity tests \\
Local fix A & Tune branch comparisons & Duplicate ambiguity tests \\
Local fix B & Skip one duplicated boundary & Some duplicate ambiguity tests \\
Correct fix & When \texttt{left == mid == right}, shrink both ends before deciding the side & None \\
\bottomrule
\end{tabular}
\end{table}

{\bf Observations.}
This example leads to four observations that directly motivate the design of our solution.

\noindent\textbf{Observation 1.}
The hard tests often expose a distinct semantic regime rather than a minor syntactic edit. In this example, the other failures require reasoning about \emph{duplicate-induced ambiguity}, not merely adjusting a comparison operator or branch~condition.

\noindent\textbf{Observation 2.}
Passing more tests does not mean that the search has reached the correct repair logic. In this example, a patch may pass most ordinary rotated-search tests yet still fail the duplicate-ambiguity, since it continues to refine the same sorted-side inference strategy instead of addressing the missing ambiguity-resolution.

\noindent\textbf{Observation 3.}
Single-trajectory refinement commits the search too early. Once methods such as REx or ChatRepair focus on one promising candidate, subsequent iterations tend to generate nearby variants of the same repair idea, reinforcing the current repair hypothesis rather than exploring alternative ones.

\noindent\textbf{Observation 4.}
Escaping the local optimum requires both diversity and abstraction shift. The search must preserve alternative repair hypotheses and be able to move from the flawed {\em ``one side is always identifiable''} strategy to the correct duplicate-aware search strategy.

\vspace{-6pt}
\section{Key Ideas}
\label{sec:ideas}

These observations motivate the key ideas of \tool. Instead of repeatedly refining a single candidate, \tool maintains a population of repair hypotheses, groups candidates by behavioral  outcomes, preserves semantically distinct repair families, and explicitly decides when search should continue exploiting the current family versus exploring a different repair abstraction. 

\begin{figure*}
    \centering
    \includegraphics[width=6.4in]{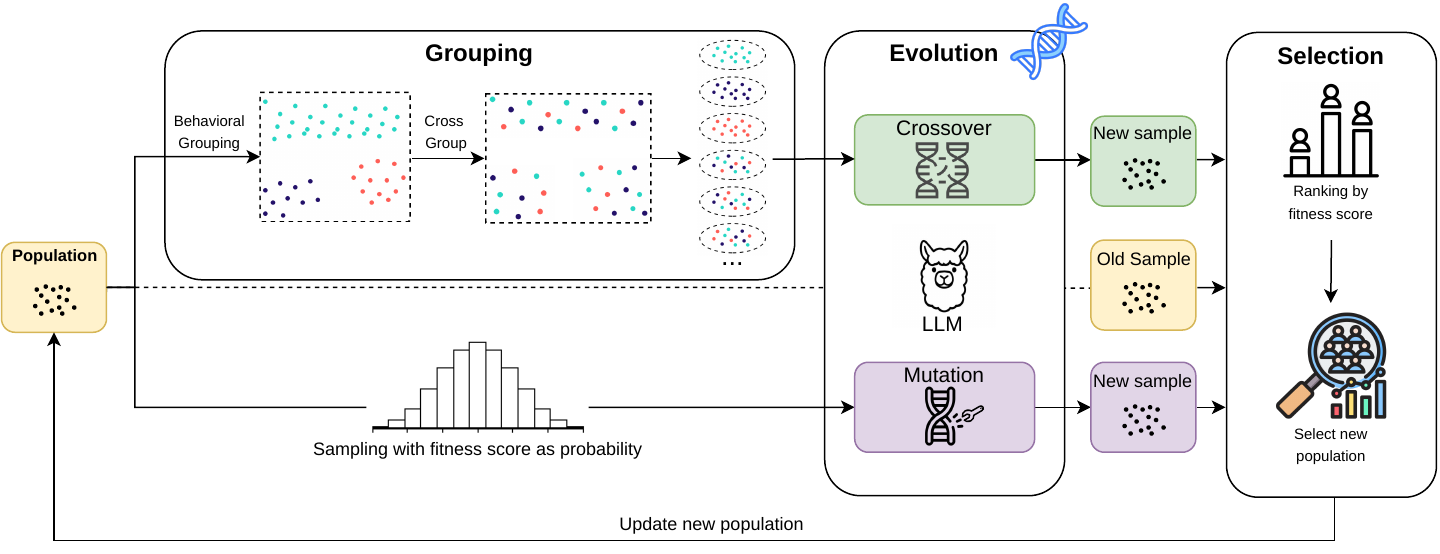}
    \vspace{-12pt}
    \caption{{\tool} Overview}
    \label{fig:overview}
\end{figure*}

\begin{enumerate} [leftmargin=0.7em, labelsep=0.2em]

\item {\bf A Population-Based Semantic Evolution formulation of LLM-driven APR.}
We {\em formulate iterative LLM-based automated program repair as a semantic evolutionary algorithm}. \tool elevates the search process from candidate-level refinement to population-level exploration, enabling systematic reasoning over diverse repair hypotheses. \tool builds on the core insight of classical evolutionary algorithms,
while replacing each syntax-driven operator with stronger, semantics-aware counterparts.

\item {\bf Semantic evolutionary operators for LLM-based repair.}
In \tool, LLMs serve as semantic mutation operators that generate context-aware edits, while behavior-aware grouping and selection preserve meaningful diversity among candidates.


\item {\bf Population-level reasoning over repair families.}
Motivated by the idea that \emph{group-relative} signals can be more informative than isolated pairwise comparisons~\cite{shao2024deepseekmath}, {\tool} lifts recombination from pairwise patch crossover to {\em population-level reasoning over repair families}. Rather than combining only two parent patches, it jointly reasons over groups of candidates with similar behavioral outcomes, enabling it to identify semantically related repairs, aggregate complementary partial fixes distributed across multiple candidates, and synthesize stronger patches that are less biased by any single parent or local repair trajectory.

\item {\bf Combining different correct partial fixes.}
Classic genetic APR uses {\em pass rate as a scalar fitness} to choose parents for crossover, but this is insufficient because it ignores {\em which} test cases a patch passes and therefore {\em what} part of the program it actually repairs. {\tool} groups candidates by their {\em passing-test subsets}, so that these execution patterns reflect which behavioral regions are already correct and which remain faulty. This enables crossover to aggregate different correct partial fixes from multiple candidates within a group, producing offspring that inherit complementary repaired behaviors and move closer to passing the full test suite.

\item {\bf Adaptive search over repair abstractions.}
By leveraging structured patterns of test failures, \tool can decide whether a repair strategy should be further refined or whether the search~should shift toward alternative repair directions, helping the algorithm escape structural local optima in iterative LLM-based~APR.

\end{enumerate}

\section{{\tool} Approach}

\subsection{Problem Definition}

We consider test-suite-based automated program repair (APR). Let $P_{\mathrm{bug}}$ denote a faulty program and let
\begin{equation}
\mathcal{T} = \{t_1, t_2, \dots, t_M\}
\end{equation}
be the associated test suite. The goal is to synthesize a repaired program $P^\star$ such that all tests pass:
\begin{equation}
\forall t \in \mathcal{T}, \quad \mathrm{Eval}(P^\star, t)=1.
\end{equation}

Equivalently, APR can be viewed as an optimization problem over the candidate repair space $\mathcal{Y}$:
\begin{equation}
P^\star \in \arg\max_{P' \in \mathcal{Y}} F(P'),
\end{equation}
where $F(P')$ measures repair quality under the test suite. In \tool, the scalar fitness is defined as the pass rate:
\begin{equation}
F(P') = \frac{1}{M}\sum_{j=1}^{M} \mathbb{I}[\mathrm{Eval}(P', t_j)=1].
\end{equation}
\tool also tracks the behavioral signature of each candidate:
\begin{equation}
\psi(P') = \{j \in \{1,\dots,M\} \mid \mathrm{Eval}(P', t_j)=1\},
\end{equation}
that is, the set of test indices passed by $P'$. This captures finer-grained behavioral information than pass rate alone, since two~candidates may obtain the same fitness while passing different tests.

\subsection{Evolutionary Formulation}

Instead of following a single repair trajectory, \tool maintains a population of candidate repairs:
\begin{equation}
\mathcal{P}^{(g)} = \{P_1^{(g)}, P_2^{(g)}, \dots, P_{N_g}^{(g)}\}, \qquad N_g \le N,
\end{equation}
where $g$ is the generation index, $N_g$ is the current population size, and $N$ is the target population size.

\tool uses a LLM as a semantic variation operator. Let $\pi_\theta$ denote the LLM parameterized by $\theta$. Given a context $c$, the LLM induces a conditional distribution over repaired programs:
\begin{equation}
P' \sim \pi_\theta(\cdot \mid c).
\end{equation}
Depending on the operator, $c$ may include the repair task, the faulty program, one or more candidate repairs, their observed test outcomes, and structured failure feedback. 

At each generation, \tool generates new candidates through recombination and mutation:
\begin{equation}
\mathcal{U}^{(g)} = \mathcal{P}^{(g)} \cup \mathcal{P}_{\mathrm{cross}}^{(g)} \cup \mathcal{P}_{\mathrm{mut}}^{(g)}.
\end{equation}
The next generation is then obtained by duplicate removal followed by top-$N$ survivor selection:
\begin{equation}
\mathcal{P}^{(g+1)} = \textsc{TopN}\big(\textsc{Unique}(\mathcal{U}^{(g)})\big).
\end{equation}
In the implementation, candidates are ranked primarily by pass rate, with ties further resolved by the number of passed tests and then by shorter code length.

This formulation highlights the central design of \tool: repair search is performed over a population of semantically evaluated candidates, rather than over isolated programs or syntax-level~edits. 


\subsection{{\tool} Algorithm}

Figure~\ref{fig:overview} and Algorithm~\ref{alg:evolrepair} summarize the overall workflow of \tool. The method begins by constructing an initial population of repair candidates with the LLM. At each generation, candidates are first grouped by execution behavior, then additional mixed groups are formed through cross-group sampling. \tool next applies population-level recombination over these candidate pools, followed by semantic mutation on individually selected parents. Finally, all candidates are merged and the top unique repairs are retained for the next generation. 
LLMs are invoked in three stages (Algorithm~\ref{alg:evolrepair}). First, they generate the initial repair population. Second, they perform population-level recombination by synthesizing a new repair from each original or mixed candidate pool. Third, they perform semantic mutation by refining individual candidates using execution feedback. In contrast, behavioral grouping, cross-group sampling, evaluation, and survivor selection are deterministic. 




\begin{algorithm}[t]
\caption{Population-Based Semantic Evolution for APR}
\label{alg:evolrepair}
\small
\begin{algorithmic}[1]
\REQUIRE Faulty program $P_{\mathrm{bug}}$, test suite $\mathcal{T}$, population size $N$, generation budget $G$
\ENSURE Best repaired program found during search

\STATE $\mathcal{P}^{(0)} \leftarrow \textsc{InitializePopulation}(P_{\mathrm{bug}}, N)$
\STATE Evaluate candidates in $\mathcal{P}^{(0)}$
\IF{a correct repair exists in $\mathcal{P}^{(0)}$}
    \RETURN it
\ENDIF

\FOR{$g = 0$ to $G-1$}
    \STATE $\mathcal{G}^{(g)} \leftarrow \textsc{BehavioralGrouping}(\mathcal{P}^{(g)})$
    \STATE $\widetilde{\mathcal{G}}^{(g)} \leftarrow \textsc{CrossGroupSampling}(\mathcal{G}^{(g)})$
    \STATE $\mathcal{P}_{\mathrm{cross}}^{(g)} \leftarrow \textsc{PopulationRecombination}(\mathcal{G}^{(g)} \cup \widetilde{\mathcal{G}}^{(g)})$
    \STATE $\mathcal{P}_{\mathrm{mut}}^{(g)} \leftarrow \textsc{SemanticMutation}(\mathcal{P}^{(g)})$
    \STATE $\mathcal{U}^{(g)} \leftarrow \mathcal{P}^{(g)} \cup \mathcal{P}_{\mathrm{cross}}^{(g)} \cup \mathcal{P}_{\mathrm{mut}}^{(g)}$
    \STATE Evaluate candidates in $\mathcal{U}^{(g)}$
    \IF{a correct repair exists in $\mathcal{U}^{(g)}$}
        \RETURN it
    \ENDIF
    \STATE $\mathcal{P}^{(g+1)} \leftarrow \textsc{SurvivorSelection}(\mathcal{U}^{(g)}, N)$
\ENDFOR

\RETURN best candidate found
\end{algorithmic}
\end{algorithm}


\subsubsection{Initial Population Construction}

\tool first constructs an initial population by repeatedly prompting the LLM with the repair task, the faulty program, and the observed error information. Each generated candidate is extracted, executed against the test suite, and assigned both a fitness score and a behavioral signature. The best valid candidates form the initial population:
\begin{equation}
\mathcal{P}^{(0)} = \textsc{TopN}(\mathcal{C}_{\mathrm{init}}),
\end{equation}
where $\mathcal{C}_{\mathrm{init}}$ denotes the set of candidates in initialization. If any candidate already achieves fitness of $1$, we terminate immediately.

\subsubsection{Behavioral Grouping}

At each generation, \tool groups candidates according to the subsets of test cases they pass. For two candidates $P_i, P_j \in \mathcal{P}^{(g)}$, similarity is defined by Jaccard overlap over their passed-test sets:
\begin{equation}
\mathrm{sim}(P_i, P_j) =
\begin{cases}
\dfrac{|\psi(P_i)\cap\psi(P_j)|}{|\psi(P_i)\cup\psi(P_j)|}, & \psi(P_i)\cup\psi(P_j)\neq\varnothing,\\[6pt]
0, & \text{otherwise}.
\end{cases}
\end{equation}
The corresponding distance is
$d(P_i, P_j) = 1 - \mathrm{sim}(P_i, P_j).$

Given a maximum allowed number of groups $K_{\max}$, we use
\begin{equation}
K_g = \max\!\left(1,\min\!\left(K_{\max}, \left\lfloor \frac{N_g}{2} \right\rfloor\right)\right)
\end{equation}
as the effective number of groups at generation $g$. Agglomerative clustering with average linkage is then applied to the precomputed distance matrix to obtain
\begin{equation}
\mathcal{G}^{(g)} = \{G_1^{(g)}, G_2^{(g)}, \dots, G_{K_g}^{(g)}\}.
\end{equation}

The implementation further rebalances singleton groups by either borrowing the closest candidate from a larger group or merging into the nearest group. As a result, each group represents a family of candidates that exhibit similar execution behavior, even when their source code differs syntactically.

\subsubsection{Cross-Group Sampling}

After behavioral groups are formed, \tool creates additional mixed groups by sampling candidates across groups. This allows recombination to combine information from different behavioral regions of the search space. To bias sampling toward behaviorally diverse groups, we reuse the similarity defined in the previous subsection. For candidates $P_a, P_b \in G_k^{(g)}$, let
\begin{equation}
q_{ab}^{(k)} = \mathrm{sim}(P_a, P_b).
\end{equation}
The diversity score of group $G_k^{(g)}$ is then
\begin{equation}
H(G_k^{(g)}) = - \sum_{a<b} q_{ab}^{(k)} \log \big(q_{ab}^{(k)} + \varepsilon \big),
\end{equation}
where $\varepsilon > 0$ is a small constant for numerical stability.

Suppose \tool creates $M_{\mathrm{mix}}$ mixed groups and each mixed group targets $E$ sampled elements. The number of candidates drawn from group $G_k^{(g)}$ is
\begin{equation}
n_k = \min\!\left(|G_k^{(g)}|,\left\lceil E \cdot \frac{H(G_k^{(g)})}{\sum_j H(G_j^{(g)})} \right\rceil\right).
\end{equation}

Each mixed group is then formed as 

\begin{equation}
\begin{aligned}
\widetilde{G}_m^{(g)} &= \bigcup_k \textsc{SampleWithoutReplacement}(G_k^{(g)}, n_k), \\
&\text{for } m=1,\dots,M_{\mathrm{mix}}.
\end{aligned}
\end{equation}

If entropy-based weighting is not informative, the implementation falls back to evenly redistributing candidates into mixed groups. Therefore, cross-group sampling explicitly constructs candidate pools that contain members from multiple behavioral groups. 


\subsubsection{Population-Level Recombination (Cross-over)}

Let

\begin{equation}
\widehat{\mathcal{G}}^{(g)} = \mathcal{G}^{(g)} \cup \widetilde{\mathcal{G}}^{(g)}
\end{equation}
denote the union of the original behavioral groups and the mixed groups produced by cross-group sampling. For each candidate pool $H \in \widehat{\mathcal{G}}^{(g)}$, \tool prompts the LLM to synthesize one improved repair:
\begin{equation}
P_H' \sim \pi_\theta(\cdot \mid x, H),
\end{equation}
where $x$ denotes the repair context and $H$ contains multiple candidate repairs together with their observed test behavior.

\begin{figure}[t]
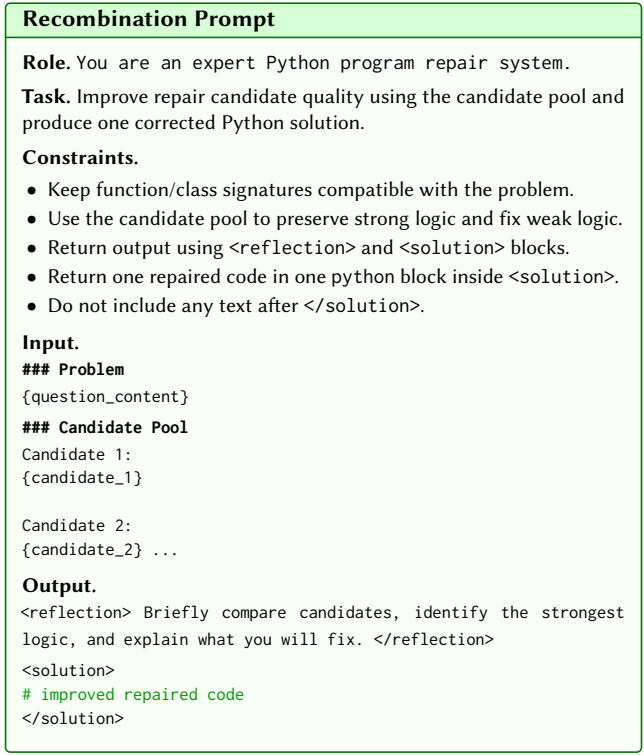

\centering
\begin{tcolorbox}[
enhanced,
width=\columnwidth,
colback=green!3!white,
colframe=green!45!black,
colbacktitle=green!16!white,
coltitle=black,
boxrule=0.6pt,
arc=1.5pt,
left=4pt,right=4pt,top=4pt,bottom=4pt,
boxsep=2pt,
title=\textbf{Recombination Prompt},
fonttitle=\bfseries\sffamily
]

\small
\sffamily
\textbf{Role.} \texttt{You are an expert Python program repair system.}

\vspace{3pt}
\textbf{Task.} Improve repair candidate quality using the candidate pool and produce one corrected Python solution.

\vspace{3pt}
\textbf{Constraints.}
\begin{itemize}
    \item Keep function/class signatures compatible with the problem.
    \item Use the candidate pool to preserve strong logic and fix weak logic.
    \item Return output using \texttt{<reflection>} and \texttt{<solution>} blocks.
    \item Return one repaired code in one \texttt{python} block inside \code{<solution>}.
    \item Do not include any text after \texttt{</solution>}.
\end{itemize}

\vspace{2pt}
\textbf{Input.}

{\footnotesize\ttfamily
\textbf{\#\#\# Problem}
}

\begin{lstlisting}
{question_content}
\end{lstlisting}

{\footnotesize\ttfamily
\textbf{\#\#\# Candidate Pool}
}

\begin{lstlisting}
Candidate 1:
{candidate_1}

Candidate 2:
{candidate_2} ...
\end{lstlisting}

\vspace{2pt}
\textbf{Output.}

{\footnotesize\ttfamily
<reflection> Briefly compare candidates, identify the strongest logic, and explain what you will fix. </reflection>
}

\vspace{2pt}
\begin{lstlisting}
<solution>
# improved repaired code
</solution>
\end{lstlisting}

\end{tcolorbox}
\vspace{-12pt}
\caption{Prompt template for recombination. The model synthesizes a stronger repair by integrating complementary logic from multiple candidate solutions.}
\label{fig:recombination-prompt}
\end{figure}

Unlike classical two-parent crossover, this operator is population-level: the LLM receives an entire candidate pool and integrates complementary logic into a single new fix. In practice, this enables \tool to combine partial fixes that arise in different behavioral groups. The prompt template for this operator is in Figure~\ref{fig:recombination-prompt}.

\subsubsection{Semantic Mutation}

\lstset{
  language=Python,
  basicstyle=\ttfamily\footnotesize,
  columns=fullflexible,
  keepspaces=true,
  showstringspaces=false,
  numbers=none,
  frame=none,
  aboveskip=2pt,
  belowskip=2pt
}

\setlist[itemize]{leftmargin=1.2em,itemsep=1pt,topsep=2pt,parsep=0pt,partopsep=0pt}

After recombination, \tool performs mutation at the individual-candidate level. Mutation parents are sampled directly from the current population with probability proportional to fitness:
\begin{equation}
\Pr(P_i^{(g)} \text{ selected}) =
\frac{F(P_i^{(g)})+\epsilon}{\sum_j \left(F(P_j^{(g)})+\epsilon\right)},
\end{equation}
where $\epsilon > 0$ ensures nonzero probability for valid candidates.

\begin{figure}[t]
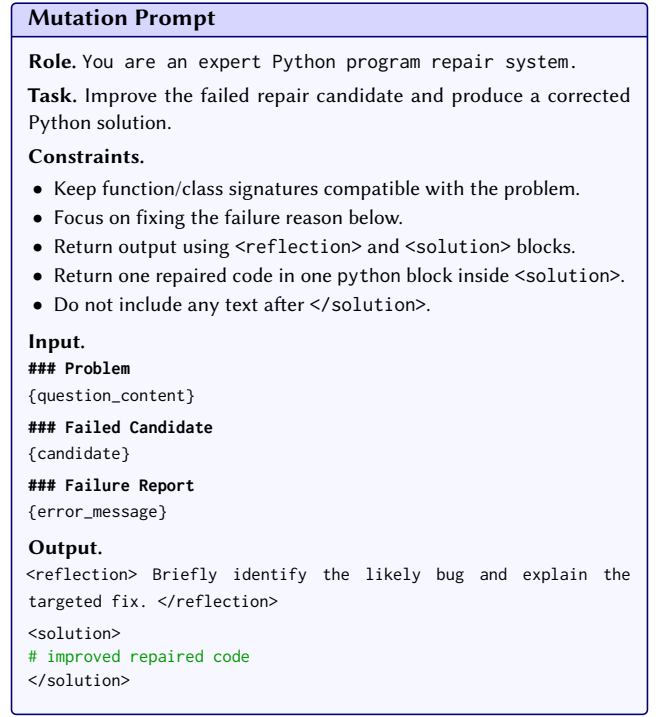

\centering
\begin{tcolorbox}[
enhanced,
width=\columnwidth,
colback=blue!3!white,
colframe=blue!45!black,
colbacktitle=blue!16!white,
coltitle=black,
boxrule=0.6pt,
arc=1.5pt,
left=4pt,right=4pt,top=4pt,bottom=4pt,
boxsep=2pt,
title=\textbf{Mutation Prompt},
fonttitle=\bfseries\sffamily
]

\small
\sffamily
\textbf{Role.} \texttt{You are an expert Python program repair system.}

\vspace{3pt}
\textbf{Task.} Improve the failed repair candidate and produce a corrected Python solution.

\vspace{3pt}
\textbf{Constraints.}
\begin{itemize}
    \item Keep function/class signatures compatible with the problem.
    \item Focus on fixing the failure reason below.
    \item Return output using \texttt{<reflection>} and \texttt{<solution>} blocks.
    \item Return one repaired code in one \texttt{python} block inside \code{<solution>}.
    \item Do not include any text after \texttt{</solution>}.
\end{itemize}

\vspace{2pt}
\textbf{Input.}

{\footnotesize\ttfamily
\textbf{\#\#\# Problem}
}

\begin{lstlisting}
{question_content}
\end{lstlisting}

{\footnotesize\ttfamily
\textbf{\#\#\# Failed Candidate}
}

\begin{lstlisting}
{candidate}
\end{lstlisting}

{\footnotesize\ttfamily
\textbf{\#\#\# Failure Report}
}

\begin{lstlisting}
{error_message}
\end{lstlisting}

\vspace{2pt}
\textbf{Output.}

{\footnotesize\ttfamily
<reflection> Briefly identify the likely bug and explain the targeted fix. </reflection>
}

\vspace{2pt}
\begin{lstlisting}
<solution>
# improved repaired code
</solution>
\end{lstlisting}

\end{tcolorbox}
\vspace{-9pt}
\caption{Prompt template for mutation. The model refines a failed candidate using structured failure feedback.}
\label{fig:mutation-prompt}
\end{figure}

For each selected parent $P_i^{(g)}$, \tool extracts a failure report $E(P_i^{(g)})$ from the execution outcome and prompts the LLM to produce a revised candidate:

\begin{equation}
P_{i,\mathrm{mut}}' \sim \pi_\theta(\cdot \mid x, P_i^{(g)}, E(P_i^{(g)})).
\end{equation}
This mutation operator is semantic: it is guided by testcase-level feedback such as failing inputs, expected outputs, actual outputs, or runtime errors. The corresponding prompt template is in Figure~\ref{fig:mutation-prompt}.

\subsubsection{Survivor Selection}

After recombination and mutation, \tool forms the expanded candidate pool
\begin{equation}
\mathcal{U}^{(g)} = \mathcal{P}^{(g)} \cup \mathcal{P}_{\mathrm{cross}}^{(g)} \cup \mathcal{P}_{\mathrm{mut}}^{(g)}.
\end{equation}
Duplicate programs are removed, and the remaining candidates are ranked to produce the next generation:
\begin{equation}
\mathcal{P}^{(g+1)} = \textsc{TopN}\big(\textsc{Unique}(\mathcal{U}^{(g)})\big).
\end{equation}
Ranking is based first on pass rate, then on the number of passed tests, and finally on shorter code length. Hence, diversity is encouraged upstream through behavioral grouping and cross-group sampling, while final survivor selection remains fitness-driven.

\section{Empirical Evaluation}
\label{sec:exp}

\subsection{Research Questions}

To evaluate {\tool}, we seek to answer the following questions.

\noindent \textbf{RQ1 [Overall Performance].}  
How effective is {\tool} in program repair compared with baseline methods?

\noindent\textbf{RQ2 [Test case Diversity Coverage].}  
Does {\tool} cover a more diverse set of bugs than baseline methods? 

\noindent\textbf{RQ3 [Cross-over Effectiveness].}  
How effective is our cross-over in combining partial fix patches? 

\noindent\textbf{RQ4 [Ablation Study].}  
How important are the key components of {\tool}? 

\noindent\textbf{RQ5 [Hyperparameter Sensitivity].}  
How sensitive is {\tool} to different hyperparameter choices, and what configurations provide reasonable default settings?

\subsection{Dataset Benchmarks \& Baselines}
\noindent \textbf{Benchmarks}. 
To reduce potential data leakage from widely used repair benchmarks, we followed SWE-Synth~\cite{pham2026swesynth}, a LLM-based bug synthesis method to build a dataset
derived from Live\-CodeBench~\cite{jain2025livecodebench}, which collects programming problems. Specifically, we randomly sampled 400 problems and generated six candidate implementations for each using Claude 3.5 Sonnet and Haiku in a 3:1 ratio to obtain solutions with diverse quality levels. Each program was executed against the provided test suite, and we retained only partially correct implementations that passed some test cases but fail at least one test case, discarding fully correct or incorrect solutions. This  yields \textbf{326 buggy solutions across 120 problems}, forming repair tasks where the goal is to modify the program to pass all test cases.

\noindent \textbf{Baselines}. 
We compare {\tool} with representative LLM-based program repair approaches which use {\em direct repair prompting} or {\em iterative refinement}. We do not compare against agentic or multi-agent APR frameworks because they operate under a different problem setting that intertwines repair quality with tool orchestration, planning, and environment interaction, whereas our focus is on {\em isolating the effectiveness of population-based search in non-agentic iterative LLM repair}. Therefore, we chose two state-of-the-art APR {\em iterative refinement} approaches.
First, \textbf{REx}~\cite{rex2024neurips} models repair as an exploration-exploitation process that iteratively samples and evaluates candidate patches to balance searching new directions and refining promising ones. \textbf{ChatRepair}~\cite{xia2024automated} adopts a conversational repair strategy in which the LLM iteratively refines patches through multi-turn interactions guided by test feedback.

\noindent \textbf{Model Backbones.} We use different LLMs such as Llama 3.3 70B Instruct, Kimi K2 Instruct 0905, and DeepSeek V3.1 as the base models, hosted via Fireworks AI for cost-efficient inference.


\vspace{-6pt}
\section{Overall Repair Performance (RQ1)}
\label{sec:rq1}

\paragraph{Experimental setting and metrics.}
We use the default hyperparameters for \tool as follows. The maximum number of evolutionary turns is 5, and the initial population contains 6 candidate patches. In each turn, behavior grouping partitions the population into 2 groups based on testcase-level passing behavior. For crossover, we further construct 2 crossing groups, each containing 3 sampled candidates, so that recombination can leverage both behaviorally similar groups and mixed groups with more diverse signals. In total, 4 group-level crossover operations are performed per evolutionary turn. In addition, we perform 3 mutation operations per turn. For \textsc{REx}~\cite{rex2024neurips} and \textsc{ChatRepair}~\cite{xia2024automated}, we set the maximum number of repair attempts to 41 to ensure a fair comparison under a comparable prompting budget. We use this configuration for later research questions (unless otherwise stated) and we will present the results with various configurations in Section~\ref{sec:rq5}.
%
We report pass@1 and pass@3 as the main repair effectiveness metrics. For each method on each buggy program, we run the repair process {\bf independently 5 times} and compute {\bf pass@k} from the generated repair outcomes. A repair is counted as \emph{correct} only if the generated program passes the \emph{entire} test suite. Partial repairs that achieve a high test pass rate but fail even a single test case are counted as failures. 
A patch that passes almost all tests but still fails one remaining case is not~useful.

\begin{table}[t]
\centering
\caption{APR performance comparison (RQ1). Correct fixes must passes all tests. Avg. RT (average runtime) and Avg. Cost (average token cost) are per buggy program. Pass@k is \%.}
\label{tab:rq1_overall}
\vspace{-6pt}
\small
\tabcolsep 2.2pt
\begin{tabular}{llcccc}
\toprule
\textbf{Backbone} & \textbf{Method} & \textbf{Pass@1} & \textbf{Pass@3} & \textbf{Avg.RT(s)} & \textbf{Avg.Cost(\$)} \\
\midrule
\multirow{4}{*}{}
& Naive Prompting & 30.98 & 33.12 & --- & --- \\
Llama 3.3 & ChatRepair~\cite{xia2024automated} & 34.36 & 40.76 & 477.96 & 0.03375 \\
 Instruct & REx~\cite{rex2024neurips} & 38.34 & 44.91 & 496.57 & 0.03015 \\
& \textbf{\tool} & \textbf{43.25} & \textbf{48.87} & 491.20 & 0.03312 \\
\midrule
\multirow{4}{*}{}
& Naive Prompting & 73.01 & 76.92 & --- & --- \\
Kimi K2 & ChatRepair~\cite{xia2024automated} & 85.92 & 88.09 & 160.27 & 0.02317 \\
 Instruct & REx~\cite{rex2024neurips} & 90.26 & 92.51 & 158.12 & 0.01821 \\
& \textbf{\tool} & \textbf{92.94} & \textbf{95.24} & 179.11 & 0.02293 \\
\midrule
\multirow{4}{*}{}
& Naive Prompting & 82.52 & 85.54 & --- & --- \\
DeepSeek & ChatRepair~\cite{xia2024automated} & 93.25 & 95.27 & 54.89 & 0.00611 \\
V3.1 & REx~\cite{rex2024neurips} & 94.48 & 95.32 & 74.07 & 0.00982 \\
& \textbf{\tool} & \textbf{96.63} & \textbf{98.18} & 77.30 & 0.01101 \\
\bottomrule
\end{tabular}
\end{table}

\vspace{1pt}
\emph{Results.}
Table~\ref{tab:rq1_overall} shows the performance,
where {\tool}, REx, and ChatRepair are evaluated as alternative iterative refinement strategies on the same three underlying models. Overall, \tool achieves the best results across both \textbf{pass@1} and \textbf{pass@3} on both backbones, showing that population-based semantic evolution is more effective than single-trajectory refinement.


Under \textbf{Llama 3.3 Instruct}, \tool achieves \textbf{43.25\%} pass@1 and \textbf{48.87\%} pass@3, outperforming \textsc{REx} (38.34\% / 44.91\%), \textsc{ChatRepair} (34.36\% / 40.76\%), and Naive Prompting (30.98\% / 33.12\%). Compared with the best baseline, \textsc{REx}, \tool improves pass@1 by 4.91 points and pass@3 by 3.96 points. These gains indicate that maintaining multiple repair hypotheses and evolving them through semantic mutation and recombination is more effective than repeatedly refining a single dominant candidate.

The same pattern holds under \textbf{Kimi K2 Instruct} and \textbf{DeepSeek V3.1}. On Kimi K2 Instruct, \tool reaches \textbf{92.94\%} pass@1 and \textbf{95.24\%} pass@3, outperforming both \textsc{REx} (90.26\% / 92.51\%) and \textsc{ChatRepair} (85.92\% / 88.09\%). On DeepSeek V3.1, \tool achieves the strongest performance, with \textbf{96.63\%} pass@1 and \textbf{98.18\%} pass@3, surpassing \textsc{REx} (94.48\% / 95.32\%), \textsc{ChatRepair} (93.25\% / 95.27\%), and Naive Prompting (82.52\% / 85.54\%). This consistent advantage across backbones suggests that {\tool}'s benefit comes from the search strategy itself rather than any LLM.

On \emph{efficiency analysis}, despite introducing population-level reasoning, behavioral grouping, and crossover, \tool remains competitive in runtime and cost. Under \textbf{Llama 3.3 Instruct}, \tool requires \textbf{491.20 seconds and \$0.03312} per buggy program on average, compared with 477.96s / \$0.03375 for \textsc{ChatRepair} and 496.57s / \$0.03015 for \textsc{REx}. The runtime is slightly higher because weaker candidates under Llama 3.3 more often fail many test cases and thus incur longer evaluation before timeout-based termination. Under \textbf{DeepSeek V3.1}, \tool requires \textbf{77.30 seconds and \$0.01101}, compared with 54.89s / \$0.00611 for \textsc{ChatRepair} and 74.07s / \$0.00982 for \textsc{REx}. Although \tool incurs slightly higher overhead on this backbone, the cost remains modest relative to the repair gains.
We do not report efficiency statistics for Naive Prompting because it consists of only a single prompting step, making it meaningless for comparing on iterative repair efficiency.

\section{Behavioral Coverage and Partial Repair Effectiveness (RQ2)}
\label{sec:rq2}

While RQ1 evaluates the final repair success of each approach, RQ2 examines how different methods behave during the search process. To this end, we analyze not only whether an approach finds a fully correct patch, but also how effectively it accumulates and consolidates partial repair progress throughout refinement.

We consider three complementary metrics. First, \textbf{Average Pass Rate (APR)} measures the average pass rate of the best candidate found for each buggy program. For each run, we select the candidate that passes the largest number of test cases and compute its pass ratio over the full test suite, then average the result across all programs and runs. APR reflects how close an approach can get to a complete repair, even when it does not produce a fully correct patch. Second, \textbf{Test Case Coverage (TCC)} measures the percentage of test cases covered by at least one generated candidate during a repair run. Unlike APR, which focuses only on the single best candidate, TCC considers the combined coverage of passed test cases across all candidates and thus captures the overall behavioral region explored during search. Third, \textbf{Coverage Gap} ($\Delta$), defined as $\Delta$ = $\mathrm{TCC}-\mathrm{APR}$, measures the gap between the capability of the best single candidate and the total behavioral coverage collectively discovered during refinement. 

Let $T_x$ be the test suite of buggy program $x$, $\mathcal{C}_{m,x,r}$ be the set of candidates generated by approach $m$ on program $x$ in run $r$, and $\psi(P)$ be the set of test cases passed by candidate $P$. Formally:
\begin{equation}
\mathrm{APR}(m) =
\frac{1}{|\mathcal{X}|R}
\sum_{x \in \mathcal{X}}
\sum_{r=1}^{R}
\max_{P \in \mathcal{C}_{m,x,r}}
\frac{|\psi(P)|}{|T_x|},
\end{equation}
\begin{equation}
\mathrm{TCC}(m) =
\frac{1}{|\mathcal{X}|R}
\sum_{x \in \mathcal{X}}
\sum_{r=1}^{R}
\frac{\left|
\bigcup_{P \in \mathcal{C}_{m,x,r}} \psi(P)
\right|}{|T_x|},
\end{equation}
\begin{equation}
\Delta(m) = \mathrm{TCC}(m) - \mathrm{APR}(m).
\end{equation}
Higher values are better for APR and TCC. For $\Delta$, a smaller value indicates that the best candidate tends to successfully maintain most of the useful repair strategies discovered during refinement.

\begin{table}[t]
\caption{Behavioral coverage and partial repair quality (RQ2).}
\label{tab:rq2_results}
\centering
\small
\vspace{-6pt}
\begin{tabular}{llccc}
\toprule
\textbf{Backbone} & \textbf{Method} & \textbf{APR (\%)}  & \textbf{TCC (\%)}  & \textbf{$\Delta$ (\%}) \\
\midrule
\multirow{4}{*}{}
& Naive Prompting   & 67.35 & --- & --- \\
Llama 3.3 & ChatRepair        & 71.85 & 74.39 & 2.54 \\
Instruct & REx               & 74.43 & 77.71 & 3.28 \\
& \textbf{\tool}   & \textbf{78.32} & \textbf{80.05} & \textbf{1.73} \\
\midrule
\multirow{4}{*}{}
& Naive Prompting   & 90.33 & --- & --- \\
Kimi K2 & ChatRepair        & 93.85 & 94.31 & 0.46 \\
Instruct & REx               & 96.27 & 97.53 & 1.26 \\
& \textbf{\tool}   & \textbf{98.33} & \textbf{98.63} & \textbf{0.30} \\
\midrule
\multirow{4}{*}{}
& Naive Prompting   & 95.57 & --- & --- \\
DeepSeek & ChatRepair        & 98.78 & 98.84 & 0.06 \\
V3.1 & REx               & 99.49 & 99.56 & 0.07 \\
& \textbf{\tool}   & \textbf{99.69} & \textbf{99.71} & \textbf{0.02} \\
\bottomrule
\end{tabular}
\end{table}

\emph{Results.}
Table~\ref{tab:rq2_results} shows that {\tool} consistently achieves the highest APR and TCC across all three backbone models, while also producing the smallest coverage gap $\Delta$. This result indicates that {\tool} not only {\em explores broader behavioral regions during repair}, but also {\em more effectively consolidates the discovered partial fixes into a better candidate}. That is, it {\em avoids getting trapped in a narrow local optimum around one evolving patch} as it can {\em preserve diverse repair hypotheses and progressively combine them, leading to candidates that are both stronger and behaviorally more complete}.

This reveals a key difference in search behavior. Iterative refinement methods mainly improve one candidate at a time, so useful partial fixes found across different candidates often remain fragmented. In contrast, {\tool} maintains a population of repair hypotheses and explicitly recombines complementary partial fixes
(Section~\ref{sec:rq3}). As a result, the best candidate in {\tool} can utilize more of the repair strategies discovered during refinement, bringing it closer to the correct patch. This result confirms our {\bf Key Idea 5}.

\begin{figure}[t]
\centering
\includegraphics[height=2.4in]{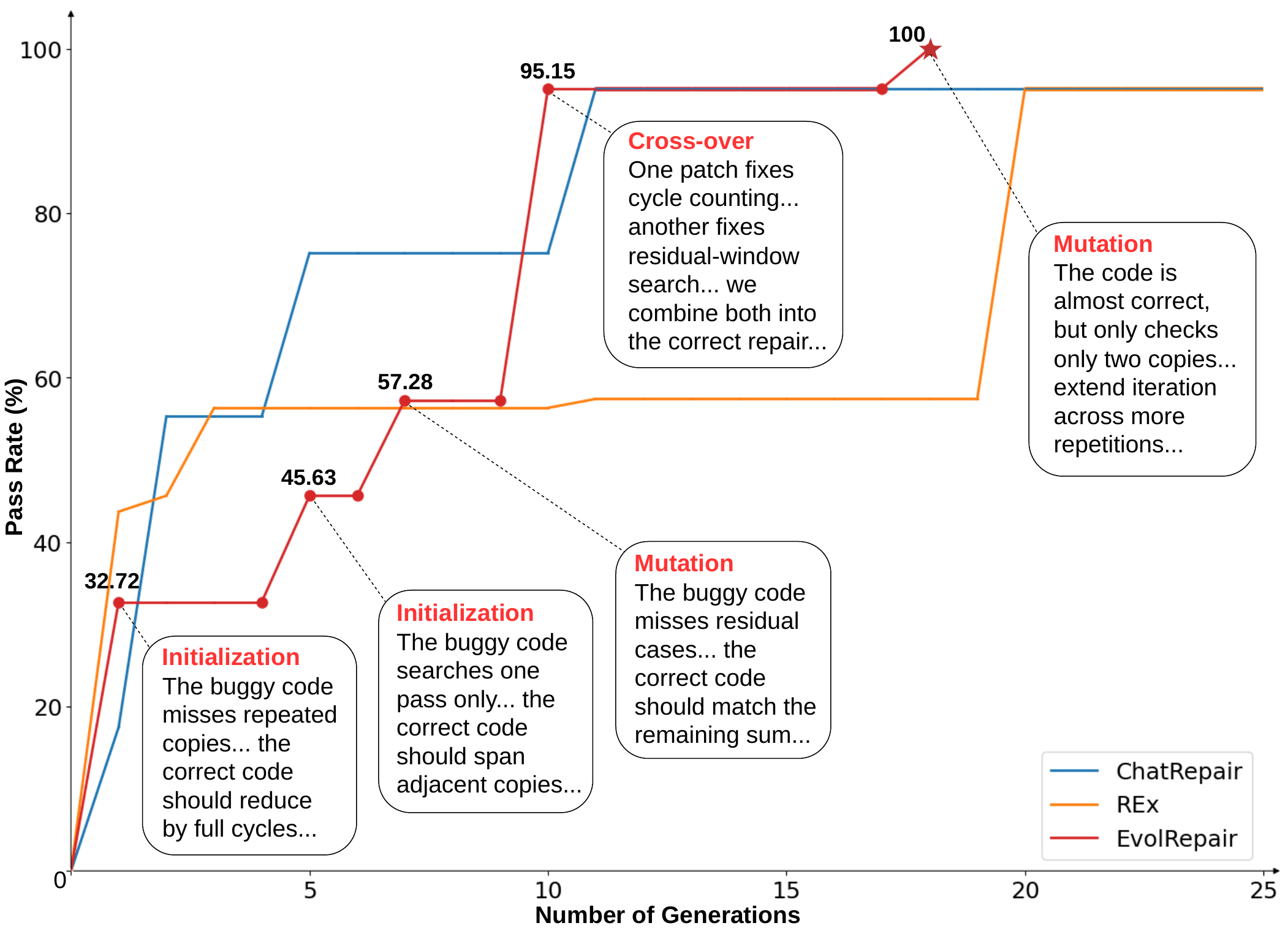} 
\vspace{-20pt}
\caption{APR progress across iterations on \texttt{mini\-Size\-Sub\-array}, which asks for the shortest target-sum subarray in an infinitely repeated array. {\tool} reaches full correctness, while baseline methods plateau at partial correctness.}
\label{fig:rq2_case}
\end{figure}

\vspace{1pt}
\emph{Case Study.}
Figure~\ref{fig:rq2_case} shows a representative example.
The baselines improve quickly~and early but plateau at a local optimum of \textbf{95.15\%}. 
This means their best repaired candidates still fail at least one test case and thus do not constitute correct repairs. Although more test cases may be covered by other intermediate candidates, these partial fixes are not effectively consolidated, resulting in a larger coverage gap. In contrast, {\tool} continues to improve over iterations and ultimately reaches \textbf{100\%}, showing its ability to escape local optima, combining partial fixes into a fully correct one.


\section{Effectiveness of Semantic Crossover for Combining Partial Repairs (RQ3)}
\label{sec:rq3}

A key design goal of {\tool} is to enable the combination of complementary partial fixes across different candidates through population-level recombination (\textbf{Key Idea 4}). In this section, we empirically evaluate whether the proposed crossover operator can synthesize new candidates that integrate correct behaviors distributed across multiple parents.

Recall that each candidate repair $P$ is associated with a behavioral signature $\psi(P)$, defined as the set of test cases it passes. During crossover, the LLM generates a new candidate $C$ from a parent candidate pool $\mathcal{P} = \{P_1, ..., P_k\}$.

\emph{Combination Metric.}
To measure whether crossover combines different complementary partial fixes, we remove test cases that are shared by two or more parents, so that each parent is associated only with its non-overlapping behavioral contribution. Let $\psi'(P_i)$ denote the remaining passing-test set of parent $P_i$ after this filtering step. If any parent has $\psi'(P_i)=\emptyset$, we exclude that~crossover group from the calculation.
We then define a binary indicator:
\begin{equation}
\text{combine}(C, \mathcal{P}) =
\begin{cases}
1, & \text{if } \forall P_i \in \mathcal{P},\ \psi(C) \cap \psi'(P_i) \neq \emptyset \\
0, & \text{otherwise}
\end{cases}
\end{equation}

This condition requires that {\em the child patch preserves at least one distinct passing test case contributed by each parent after removing shared behaviors}. In other words, a crossover is counted as successful only when the child covers at least one unique behavioral contribution from every parent in the group. We then define the overall \textbf{Combination Rate} as:
\begin{equation}
\text{Combination Rate} =
\frac{\sum \text{combine}(C, \mathcal{P})}{\text{total number of cross-over operations}}
\end{equation}

\paragraph{Results.}
Table~\ref{tab:crossover_effectiveness} reports the combination rate across different backbone models. As seen, semantic crossover can effectively combine complementary partial fixes across candidates. In particular, the non-trivial combination rates across all three backbones indicate that the generated offspring often preserve distinct behavioral contributions from multiple parents, rather than merely reproducing or slightly modifying a single repair trajectory.

This provides direct evidence that {\tool}'s population-level recombination operator is genuinely compositional: it can merge previously separated correct behaviors into one stronger candidate. The especially high rate on Llama 3.3 Instruct indicates that, even when the backbone is comparatively weaker, crossover can still effectively aggregate distributed repair evidence, thereby contributing substantially to the superior repair performance over single-trajectory refinement baselines. Overall, these results support {\bf Key Idea 4}: population-level recombination over behaviorally grouped candidates enables effective composition of partial repairs.

\begin{table}[t]
\centering
\small
\caption{Effectiveness of semantic cross-over (RQ3).}
\label{tab:crossover_effectiveness}
\vspace{-6pt}
\begin{tabular}{l c}
\toprule
\textbf{Backbone Model} & \textbf{Combination Rate (\%)} \\
\midrule
Llama 3.3 Instruct & 61.72 \\
Kimi K2 Instruct & 19.51 \\
DeepSeek V3.1 & 35.71 \\
\bottomrule
\end{tabular}
\end{table}

\begin{figure*}[t]
\centering
\includegraphics[width=0.72\linewidth]{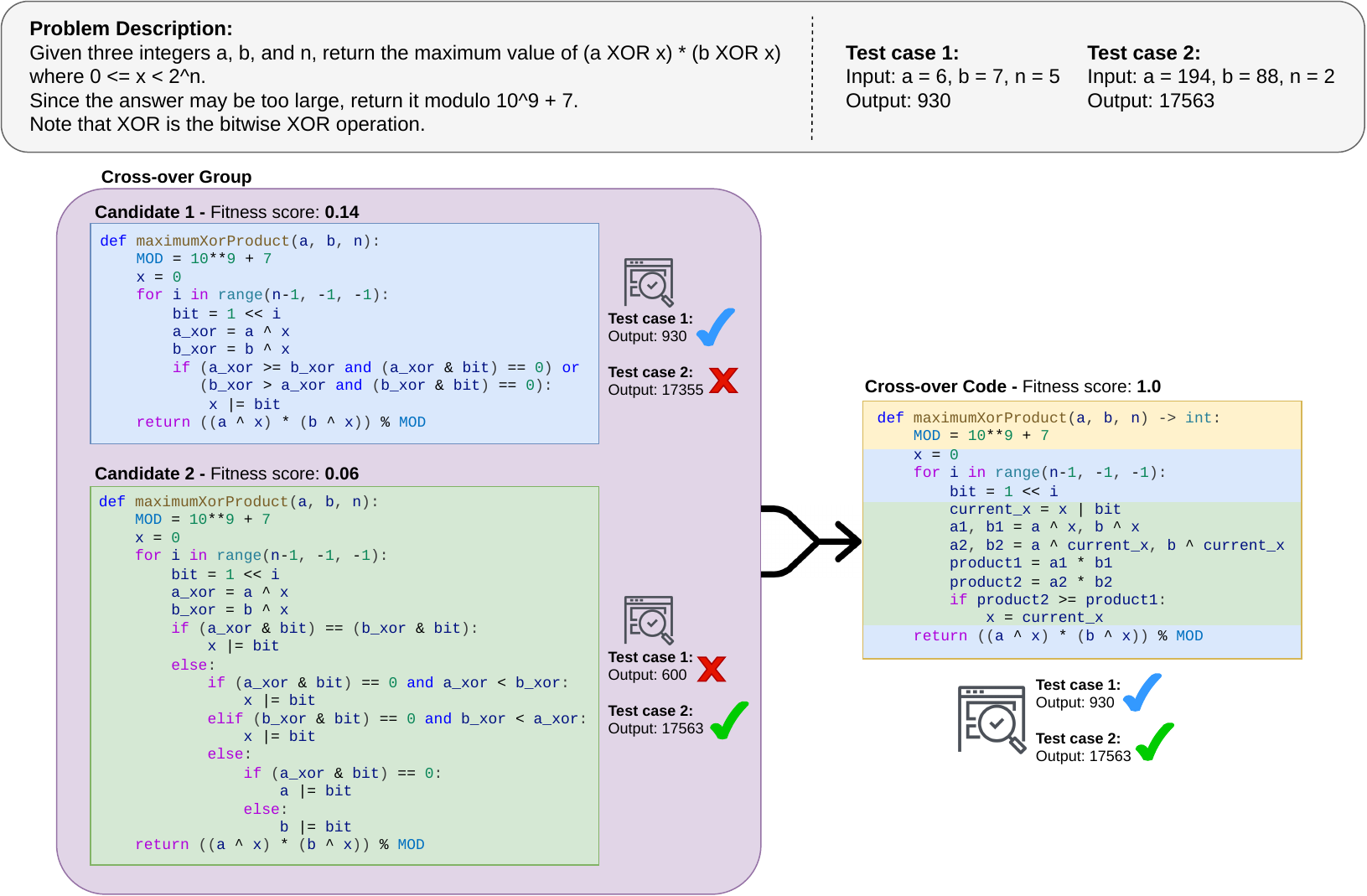}
\vspace{-9pt}
\caption{Example of semantic crossover combining complementary partial fixes into a correct repair that passes all testcases; \textcolor{blue}{blue} parts of the cross-over code are derived from Candidate~1, and \textcolor{green!50!black}{green} parts are derived from Candidate~2.}
\label{fig:crossover_case}
\end{figure*}

\vspace{1pt}
\emph{Case Study.}
Figure~\ref{fig:crossover_case} illustrates an example where the child integrates complementary fixes from multiple parents. Each parent repairs a different subset of failing test cases, while the child combines these partial fixes and achieves improved overall correctness. Specifically, the crossover preserves the greedy bitwise construction of \texttt{x} from Candidate~1, while incorporating Candidate~2’s insight of evaluating the effect of each bit choice, replacing both parents’ flawed heuristics with a direct objective-level comparison.

\section{Ablation Study (RQ4)}
\label{sec:rq4}


We evaluate five ablated variants: \textbf{Pairwise Recombination}, which replaces behavior-based grouping with classic candidate pairing before recombination; \textbf{Fitness-based Grouping}, which groups candidates according to similar fitness scores instead of passed-test behavior; \textbf{Random Grouping}, which forms groups randomly; \textbf{w/o Crossover}, which removes the semantic recombination operator; and \textbf{w/o Mutation}, which removes the semantic mutation operator. We select \textbf{DeepSeek V3.1} as the backbone model for both RQ4 and RQ5, as it provides the best overall performance with lower cost.

\begin{table}[t]
\caption{Ablation results (RQ4)}
\label{tab:rq4_ablation}
\centering
\small
\vspace{-6pt}
\begin{tabular}{lcc}
\toprule
\textbf{Setting} & \textbf{Pass@1 (\%)} & \textbf{Pass@3 (\%)} \\
\midrule
\textbf{Full EvolRepair}           & \textbf{96.63} & \textbf{98.18} \\
Pairwise Recombination    & 93.67 & 95.92 \\
Fitness-based Grouping    & 96.03 & 97.71 \\
Random Grouping           & 93.67 & 95.63 \\
w/o Crossover             & 94.50 & 96.84 \\
w/o Mutation              & 88.11 & 91.83 \\
\bottomrule
\end{tabular}
\end{table}

Table~\ref{tab:rq4_ablation} shows that {\tool} achieves the best results among all variants.
Replacing behavior-based grouping with weaker alternatives consistently reduces performance. In particular, fitness-based grouping causes only a small drop, suggesting that fitness provides a useful coarse signal, but it still underperforms behavior-based grouping because candidates with similar pass rates may repair different subsets of test cases. Random grouping performs worse, as it ignores these behavioral relations altogether. This confirms {\bf Key Idea~4}: grouping candidates by passed-test behavior better exposes complementary partial fixes for recombination.

Replacing group-based recombination with pairwise version degrades performance. This suggests that group-level recombination provides richer context than isolated candidate pairs, allowing the LLM to infer a stronger repair strategy from multiple related, complementary candidates. This supports our {\bf Key Idea~3} of recombination on behavioral groups rather than standard pairwise~crossover.

Removing crossover and mutation both hurts performance, confirming that the two operators play complementary roles. The larger drop without mutation, which gives the worst results overall, highlights the importance of test-feedback-guided refinement for improving individual candidates. The drop without crossover shows that recombination is also necessary for combining useful behaviors across repair hypotheses. Together, these results support our core design: behavior-aware grouping identifies complementary repairs, crossover composes them, and mutation refines them.

\vspace{-3pt}
\section{Sensitivity to Hyperparameter Settings (RQ5)}
\label{sec:rq5}

\begin{table}[t]
\caption{Sensitivity analysis of EvolRepair (RQ5)}
\label{tab:rq5_hyper}
\small
\centering
\vspace{-9pt}
\begin{tabular}{ccc}
\toprule
\textbf{$(\alpha,\beta,\gamma,\delta,\epsilon,\eta)$} & \textbf{Pass@1 (\%)} & \textbf{Pass@3 (\%)} \\
\midrule
$(5,6,2,2,3,3)$ & 96.63 & 98.18 \\
$(5,4,2,2,5,3)$ & 95.84 & 97.62 \\
$(4,6,3,3,3,3)$ & 95.91 & 97.88 \\
$(7,6,2,1,3,2)$ & 96.32 & 98.04 \\
$(5,6,2,3,3,2)$ & 96.11 & 98.18 \\
$(5,6,2,1,3,4)$ & 96.05 & 97.81 \\
\bottomrule
\end{tabular}
\end{table}

In this experiment, we further study sensitivity by varying one parameter at a time while keeping others fixed. The examined parameters include maximum evolutionary turns ($\alpha$), initial population size ($\beta$), number of behavior groups ($\gamma$), number of crossing groups ($\delta$), candidates per crossing group ($\epsilon$), and mutation operations per turn ($\eta$). In each turn, the population is partitioned into $\gamma$ behavior groups based on test-level behavior, followed by $\delta$ crossing groups with $\epsilon$ candidates each, enabling recombination across both similar and diverse candidates. Mutation is then applied $\eta$ times.

Our default setting $(5,6,2,2,3,3)$ corresponds to a prompting budget of approximately 40 interactions. To ensure fair comparison, we adjust configurations to keep the overall budget comparable.

Table~\ref{tab:rq5_hyper} shows that {\tool} remains robust under moderate hyperparameter variations, with only minor performance fluctuations (within $\sim$1\%). Increasing exploration (e.g., larger $\alpha$) or mutation depth ($\eta$) yields comparable results, while reducing population size ($\beta$) or grouping granularity ($\gamma,\delta$) slightly degrades performance due to lower diversity or fewer recombination opportunities.

Overall, these indicate that {\tool} is not overly sensitive to hyperparameter choices, and the default configuration provides a strong balance between exploration, recombination, and efficiency.

\vspace{-8pt}
\section{Threats to Validity}

\emph{Internal Validity.}
Results depend on parameters, e.g. behavioral grouping, selection, prompting, etc.
Although all tools use the same backbones and protocol,
such differences 
may affect performance. 

\emph{External Validity.}
We used synthetic benchmarks and a limited set of models, which may not fully represent real-world scenarios. However, it reduces the data leakage issue. Results may not generalize to other languages, large systems, or domain bugs.

\emph{Construct Validity.}
Test-suite pass rates may not fully capture correctness due to its incompleteness and may lead to overfitting. Behavioral signatures provide richer signals but are derived from test outcomes, and pass rate does not fully capture patch~quality.

\section{Related Work}
\label{sec:related}

{\bf Genetic Algorithm} (GA) has been applied to program repair.
GenProg~\cite{weimer2009genprog,le2011genprog} and jGenProg~\cite{weimer2009genprog}
realized this via genetic programming over syntactic edits.
\tool retains the population-based search intuition of classical GA-based APR systems, but upgrades each core operator with semantics-aware counterparts as explained. 

{\bf \emph{Traditional APR.}}
Early APR systems relied on search-based or template-driven techniques.
Representative approaches include 
mutation-based repair such as \textsc{jMutRepair} and \textsc{PraPR}~\cite{pacheco2019prapr},
and template-based systems (\textsc{TBar}~\cite{liu2019tbar},
\textsc{SemFix}~\cite{nguyen2013semfix}, and \textsc{AVATAR}~\cite{liu2019avatar}).
Other systems such as \textsc{CapGen}~\cite{wen2018capgen},
\textsc{SketchFix}~\cite{mechtaev2015sketchfix}, and
\textsc{JAID}~\cite{chen2021jaid} further explored constraint solving,
fault localization, or runtime analysis to synthesize patches.
Mining approaches (\textsc{FixMiner}~\cite{koyuncu2018fixminer})
leverage recurring fix patterns extracted from historical patches.

{\bf \emph{Deep-learning-based APR.}}
To expand the repair search space beyond handcrafted templates,
subsequent work leveraged neural models to learn repair patterns from
large bug-fix datasets. Systems such as \textsc{CURE}~\cite{ye2021cure}, \textsc{CoCoNuT}~\cite{lutellier2020coconut},
\textsc{Recoder}~\cite{zhu2021recoder}, and \textsc{SelfAPR}~\cite{ye2023selfapr}
frame APR as a sequence-to-sequence or edit-generation problem.
Recent models such as \textsc{Tare}~\cite{tang2023tare} and
\textsc{KNOD}~\cite{li2023knod} incorporate richer structural
representations of code to improve generation accuracy.

{\bf \emph{LLM-based APR with prompting.}}
Recent advances in large language models (LLMs) have transformed APR
by enabling powerful zero-shot and few-shot patch generation.
Studies such as Fan et al.~\cite{fan2023apr_llm_outputs} and
Xia et al.~\cite{xia2023plm_apr} show that large pretrained models can
generate plausible fixes when prompted with buggy code and context.
Several systems adopt direct prompting workflows.
For example, \textsc{AlphaRepair}~\cite{xia2022alpharepair} leverages
large pretrained code models to generate patches with infilling-style prediction, 
while \textsc{Repilot}~\cite{kang2023repilot}
integrates LLM-based code completion with patch validation.

{\bf \emph{Iterative LLM-based repair.}}
Recent work explores iterative refinement using interaction with LLMs. 
\textsc{ChatRepair}~\cite{xia2024automated}
uses conversational prompting with ChatGPT to iteratively improve
patches using test feedback and execution results.
REx~\cite{rex2024neurips} improves APR for LLMs by iterative refinement.
It formulates iterative code repair as a budgeted search problem.
In comparison, \tool shifts the decision unit from individual candidates to \emph{behavioral regimes}. 


{\bf \emph{Retrieval-augmented APR.}}
Retrieval-augmented
approaches incorporate similar bug-fix exemplars into the prompt.
For example, RAP-Gen~\cite{wang2023rapgen} retrieves relevant fixes
from historical repositories and integrates them into the LLM context
to guide patch generation.

{\bf \emph{Agentic APR frameworks.}}
\textsc{AutoCodeRover}~\cite{zhang2024autocoderover}
combines LLM reasoning with repository navigation and test execution
to autonomously repair bugs, while \textsc{RepairAgent}~\cite{bouzenia2025repairagent}
uses an LLM-driven agent to localize faults, analyze program behavior,
and iteratively repair code using external tools.

\vspace{-3pt}
\section{Conclusion}

We present {\tool}, a population-based semantic evolution framework for LLM-driven APR. By enabling diversity preservation, behavioral grouping, and composition of partial repairs, {\tool} improves over iterative refinement baselines such as REx and ChatRepair across multiple models. Our results suggest that elevating repair from single-trajectory refinement to population-level search is key to advancing LLM-based program repair.

\section{Data
Availability Statement}

All data and code are available at~\cite{anonymous_2026_19337682}

\balance

\bibliographystyle{ACM-Reference-Format}

\bibliography{references,apr-references}

\end{document}